\documentclass{aa}
\usepackage{graphicx}
\usepackage{natbib}
\bibpunct{(}{)}{;}{a}{}{,}

\begin{document}

\titlerunning{Probing MACHOs with the 1.5m Loiano telescope}
\title{Probing MACHOs by observation of M31 pixel lensing\\
    with the 1.5m Loiano telescope}

\authorrunning{Calchi Novati et al.}
\author{S.~Calchi Novati\inst{1,2} 
\and G.~Covone\inst{3}
\and F.~De Paolis\inst{4}
\and M.~Dominik\thanks{Royal Society University Research Fellow}\inst{5}
\and Y.~Giraud-H{\'e}raud\inst{6}
\and G.~Ingrosso\inst{4} 
\and Ph.~Jetzer\inst{7} 
\and L.~Mancini \inst{1,2} 
\and A.~Nucita \inst{4} 
\and G.~Scarpetta\inst{1,2}
\and F.~Strafella\inst{4} 
\and A.~Gould\inst{8}\\
(the PLAN\thanks{Pixel Lensing Andromeda} collaboration)}

\institute{
Dipartimento di Fisica ``E. R. Caianiello'', 
Universit\`a di Salerno, Via S. Allende, 84081 Baronissi (SA), Italy \and
Istituto Nazionale di Fisica Nucleare, sezione di Napoli, Italy\and
INAF - Osservatorio Astronomico di Capodimonte, via Moiariello 16, Napoli, Italy\and
Dipartimento di Fisica, Universit\`a di Lecce and INFN, Sezione
di Lecce, CP 193, 73100 Lecce, Italy\and
SUPA, University of St Andrews, School of Physics \& Astronomy, North
Haugh, St Andrews, KY16 9SS, United Kingdom\and
APC, 10, rue Alice Domon et L\'eonie Duquet 75205 Paris, France\and
Institute for Theoretical Physics,  University of Z\"urich, 
Winterthurerstrasse 190, 8057 Z\"urich, Switzerland \and
Department of Astronomy, Ohio State University, 140 West 18th Avenue, Columbus, OH 43210, US
}
\date{Received/ Accepted}

\abstract{
We analyse a series of pilot observations in order to study microlensing of
(unresolved) stars in M31 with the 1.5m Loiano telescope, including
observations on both identified variable source stars and
reported microlensing events. We also look for
previously unknown variability and discover a nova. We discuss an
observing strategy for an extended campaign with the goal of determining
whether MACHOs exist or whether all microlensing events are compatible 
with lens stars in M31.
}
\maketitle

\section{Introduction} 
Following the original proposal
of \cite{pacz86}, several microlensing campaigns have been undertaken
in the recent years with the purpose
of unveiling the content of galactic halos in form of MACHOs. 
While both the MACHO \citep{macho00} and EROS \citep{eros06} groups
have published comprehensive results of their respective campaigns,
and an analysis of the OGLE campaign is underway, no consensus has yet
been reached on either the density of MACHOs or their mass spectrum, 
and it is still not clear whether ``self lensing'' within the Magellanic Clouds
\citep{sahu94,wu94} can account for most or even all of the detected microlensing
events \citep{bel03,bel04,mancini04,griest05,bennett05,novati06,evans06}.

Searching for microlensing events towards the Andromeda Galaxy (M31) 
not only allows one to monitor a huge number of stars ($\sim 10^{8}$) within 
a few fields, but also allows one to fully probe M31's whole halo
(which is not possible for the Milky Way), and possibly
to distinguish more easily between self lensing and lensing by MACHOs
because M31's tilt with respect to the line of sight induces a 
characteristic signature in the spatial distribution of the halo events
\citep{crotts92,agape93,jetzer94}.
Observational campaigns have been carried out
by several collaborations: AGAPE \citep{agape97,agape99}, Columbia-VATT 
\citep{crotts96}, POINT-AGAPE \citep{point01,paulin03}, SLOTT-AGAPE \citep{novati02,novati03},
WeCAPP \citep{wecapp03}, MEGA \citep{mega04}, 
NainiTal \citep{nainital05} and ANGSTROM \citep{kerins06}.
The detection of a handful of microlensing candidates have
been reported and first, though contradictory, conclusions on the MACHO content
along this line of sight have been reported \citep{novati05,mega06}.

In order to go beyond these first results, it is essential to
choose an appropriate observational strategy for the new
observational campaigns. Indeed, the experience of the previous 
campaigns shows that a careful assessment of the characteristics 
of the microlensing signal and of potentially
contaminating stellar variables is crucial.
Two main phenomenological characteristics 
of microlensing events must be taken into account:
the duration and the flux deviation (e.g. \citealt{gin06a,gin06b}). 
Microlensing events in M31 are expected to last only a few days 
(this holds  in the lens mass range $10^{-2}-1$ M$_\odot$, 
over which self lensing but also most of the MACHO signal is expected).
Note that here we refer to the full-duration-at-half-maximum, $t_{1/2}$,
easily evaluated out of pixel lensing observations,
with $t_{1/2}=t_{1/2}(t_\textrm{E},u_0)$,
where $t_\textrm{E},\,u_0$ are the \emph{Einstein time} and impact parameter,
respectively. The degeneracy in the parameter space
$t_\textrm{E},u_0$ is intrinsically linked
to the fact that the underlying sources are not resolved objects
so that the background level of the light curves is a blend
(of a huge number) of stars \citep{gould96}.
However, as was shown to be the case for some of the POINT-AGAPE
microlensing candidates, extremely good sampling along the flux
variation sometimes allow one to break this degeneracy.
To gain insight into the underlying mass spectrum of the lens population
(recall $t_\textrm{E}\propto \sqrt{M_\textrm{lens}}$), it will
be essential to break the $t_\textrm{E},u_0$ degeneracy
beyond what was achieved in previous campaigns.
Furthermore, the expected short duration can also be used to robustly test the detected
flux variations with respect to the variable star background \citep{novati05},
but to achieve this, a very tight and regular sampling is again necessary.
On the other hand, the expected duration implies that
to characterise the microlensing signals,
the campaign does not need to be extremely long.
Besides, the dataset of previous campaigns already allows one
to check for the expected uniqueness of microlensing signals.
The long time baseline can then be exploited in order to increase
the expected rate of events. Very tight and regular sampling on a
nightly basis is therefore a first crucial feature for an optimal
observational strategy. This would represent an important
improvement with respect to previous campaigns
that would allow one both to better distinguish
microlensing events from other background variations,
and, possibly, to break some of the degeneracy
in the microlensing parameter space.
As for the flux deviation, the main results have
been obtained using the 2.5m INT telescope with 
integration times of about 20 minutes per night,
so that  even smaller telescopes can be used,
provided that long enough integration times are employed
to reach the needed signal-to-noise ratio.

In this paper we present the results of the pilot season 
of a new observational campaign towards M31 carried out with the 
Loiano telescope at the ``Osservatorio Astronomico 
di Bologna'' (OAB)\footnote{http://www.bo.astro.it/loiano/index.htm}.
In Sect. \ref{sec:data} we present the observational setup and outline
data reduction and analysis. In Sect. \ref{sec:res}
we present the results of our follow-up observations on previously
reported  microlensing candidates and other variable light curves,
and we report the discovery of a new Nova variation. In Sect. \ref{sec:ml}
we estimate the expected microlensing
signal and discuss the feasibility and objectives
of a longer-term microlensing campaign.

\section{Data analysis} \label{sec:data}

\subsection{Observational setup, data acquisition and reduction}

\begin{figure}
\resizebox{\hsize}{!}{\includegraphics{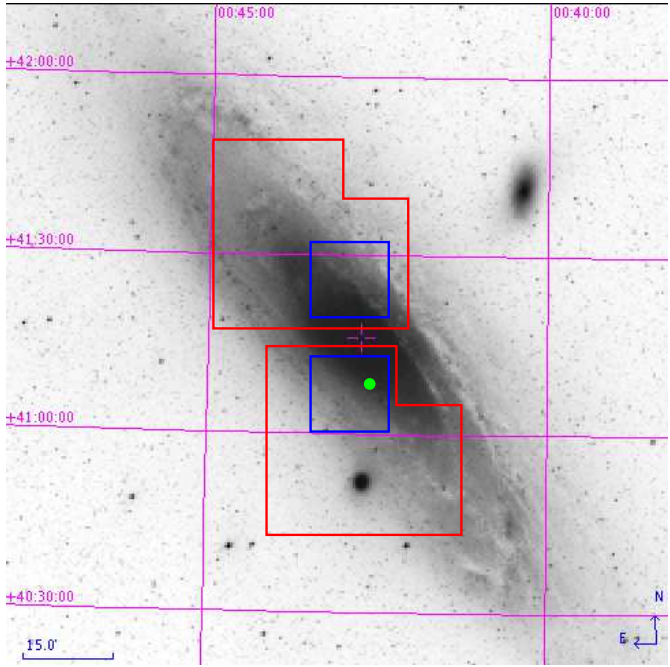}} \caption{
Projected on M31, we display the boundaries of the two 13'x12.6'
monitored fields (inner contours), together with the larger INT fields
and the centre M31 (cross). The filled circle marks the
position of the Nova variable detected (Sect. \ref{sec:nova}).} 
\label{fig:field}
\end{figure}

As pilot observations for studying microlensing of stars in the inner M31
region, we observed two fields during 11 consecutive nights, from
5 September to 15 September 2006, with the 152cm Cassini Telescope
located in Loiano (Bologna, Italy).
We make use of a CCD EEV of 1300x1340 pixels of $0.\hskip-2pt''58$
for a total field of view of $13'\times 12.\hskip-2pt'6$,
with gain of 2 e$^-$/ADU and low read-out noise (3.5 e$^-$/px).
Two fields of view around the inner  M31 region have been monitored,
centered respectively at
RA=$0{\textrm h}42{\textrm m}50{\textrm s}$, 
DEC=$41^\circ 23' 57''$ (``North'')
and RA=$0{\textrm h}42{\textrm m}50{\textrm s}$, DEC=$41^\circ 08' 23''$ (``South'') (J2000),
so as to leave out the innermost ($\sim 3'$) M31 bulge region,
and with the CCD axes parallel to the north-south and east-west directions,
in order to get the maximum field overlap with previous campaigns (Fig.~\ref{fig:field}).
To test for achromaticity, data have 
been acquired in two bandpasses (similar to Cousins $R$ and $I$),
with exposure times of 5 or 6 minutes per frame.
Overall we collected about 100 exposures per field
per filter over 8 nights or about
15 images per night\footnote{During the last useful night
only a few $R$ images in the North field could
be taken.}. Typical seeing values were in the range $1.5''-2''$.
Bias and sky flats frames were taken each night
and standard data reduction was carried out using the 
IRAF package\footnote{http://iraf.noao.edu/}.
We also corrected the $I$-band data for fringe effects.

\subsection{Image analysis}

The search for flux variations towards M31 has to deal
with the fact that sources are not resolved objects,
so that one has to monitor flux variations
of every element of the image (the so called ``pixel-lensing''
technique discussed in \citealt{gould96}). 
As for the preliminary image analysis,
we follow closely the strategy outlined by the AGAPE group
\citep{agape97,novati02}, in which each image
is geometrically and photometrically aligned
relative to a reference image.
To account for seeing variations we then substitute for
the flux of each pixel, the flux of the corresponding 
5-pixel square ``superpixel'' centered on it 
(whose size is determined so as to cover most of the average seeing disc)
and then apply an empirical, linear, correction in the flux,
again calibrating each image with respect to the reference image.
The final expression for the flux error accounts both
for the statistical error in the flux count
and for the residual error linked to the seeing correction procedure.
Finally, in order to increase the signal-to-noise ratio,
we combine the images taken during the same night.

\section{Light curve results} \label{sec:res}

\subsection{Variables in the POINT-AGAPE catalogue} \label{sec:var}

In order to assess the quality of the present data set 
as compared to that of previous campaigns
we looked into observations of $\sim$ 40000 stars
identified as variables by the POINT-AGAPE group \citep{an04}.
Besides the position, each variation in this catalogue
is characterised by three quantities: the 
magnitude corresponding to the flux deviation at maximum
$R(\Delta\phi)$, with values down to 
$R(\Delta\phi)\sim 23$, the period ($P$) as evaluated using
a Lomb algorithm, and an estimator
of the probability of a false detection ($L_f$)
(high absolute values of $L_f$ indicate a sure identification).
We note that most of the variations in the catalogue are rather faint 
and only a few have short periods.

\begin{figure}
\resizebox{\hsize}{!}{\includegraphics{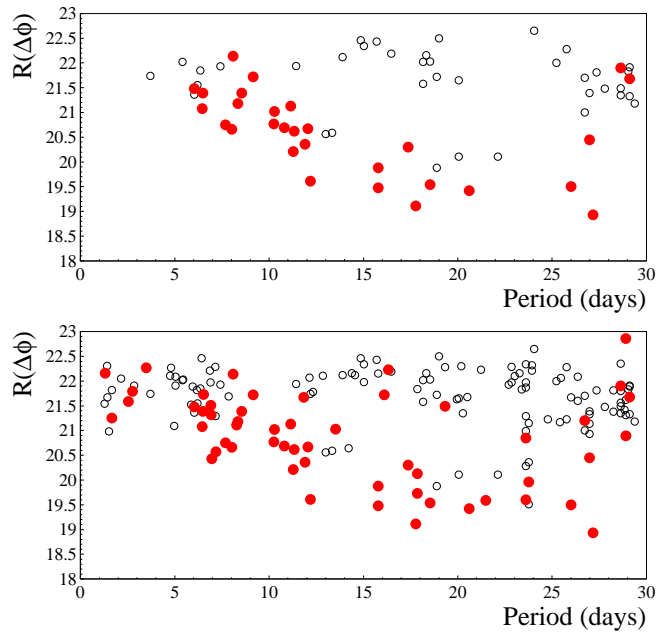}} \caption{
$R(\Delta\phi)$ vs.\ Period for short-period ($P < 30~\mbox{d}$) 
variable stars reported in the POINT-AGAPE catalogue
(top panel shows the subset with $\log(L_f)<-30$),
where filled circles indicate the flux variations identified
within the OAB data set. 
\label{fig:var-param}}
\end{figure}

We want to investigate which fraction
of variables found by POINT-AGAPE can be identified by our observations.
(Preliminary to the analysis, we must evaluate
the relative geometrical and photometrical
transformation between the two data sets. In particular,
we find that $\sim 30\%$ of the original sample
belongs also to our field of view).
Since our observations cover only 11 days, we restrict our attention to
the shortest periods ($P < 30~\mbox{d}$), which encompasses
a $\sim 2\%$ subset of the POINT-AGAPE catalogue.
Note that our limited baseline 
does not allow us to properly characterise the shape parameters
of the detected variations. Therefore,
in order to cross-identify the flux variations detected
with those belonging to the POINT-AGAPE
catalogue we only test for the offset
between the position evaluated through our selection
and the transformed POINT-AGAPE position.

\begin{figure}
\resizebox{\hsize}{!}{\includegraphics{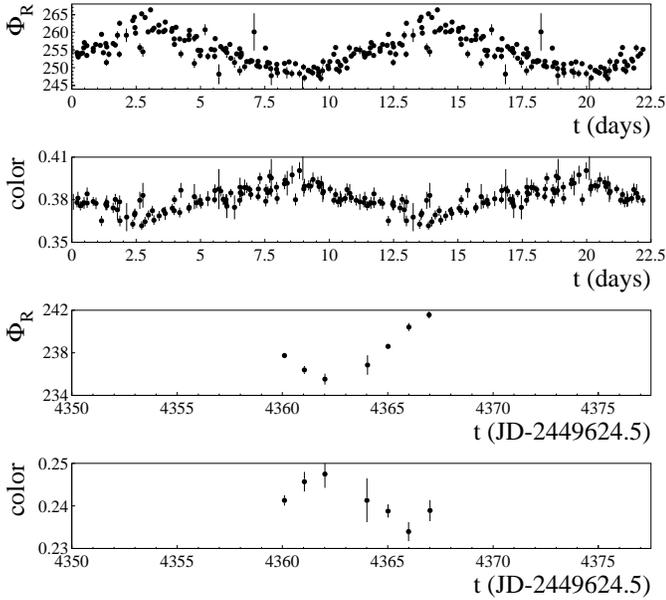}} \caption{
The light curve of a POINT-AGAPE flux variation
($P=11.14$ days and $R(\Delta\phi)=21.1$),
together with its extension in the OAB data.
Top to bottom, the INT $R$ and color light curves, folded by their period
(for visual aid, two cycles are plotted); and the OAB $R$ and color light curves.
The ``color'' is evaluated as $-2.5 \log(\phi_r/\phi_i)$,
where $\phi$ is the observed flux. Flux is in ADU s$^{-1}$.
\label{fig:var}}
\end{figure}

For our analysis, we first identify a ``clean''
set of variable stars (selected by demanding $\log(L_f)<-30$),
which includes $\sim 25\%$ of the POINT-AGAPE sample.
Restricting ourselves to short-period variables with $P~<~30~\mbox{d}$,
leaves us with 169 stars within our field of view, among which 68
fall into the ``clean'' sample.
For the latter, we detect most of the bright variations ($R(\Delta\phi)<21$),
namely $\sim 70\%$, and about 40\% of all of the variations.
When we consider the total sample of short-period variables
we arrive at values that are about 10\% smaller.
This partly results from the fact that the total sample contains a larger
fraction of faint objects, while our detection threshold, 
though varying with the position in the fields,
is typically about $R(\Delta\phi)\sim 22$. 
In Fig.~\ref{fig:var-param} we show
flux deviation vs.\ period for the
the full set of short-period POINT-AGAPE variables, where 
solid circles mark those that were found by our analysis.
In Fig.~\ref{fig:var}, we show the lightcurve of a POINT-AGAPE variable
recovered within the OAB data, with its OAB extension.

\subsection{Identified microlensing candidates} \label{sec:evts}

\begin{figure}
\resizebox{\hsize}{!}{\includegraphics{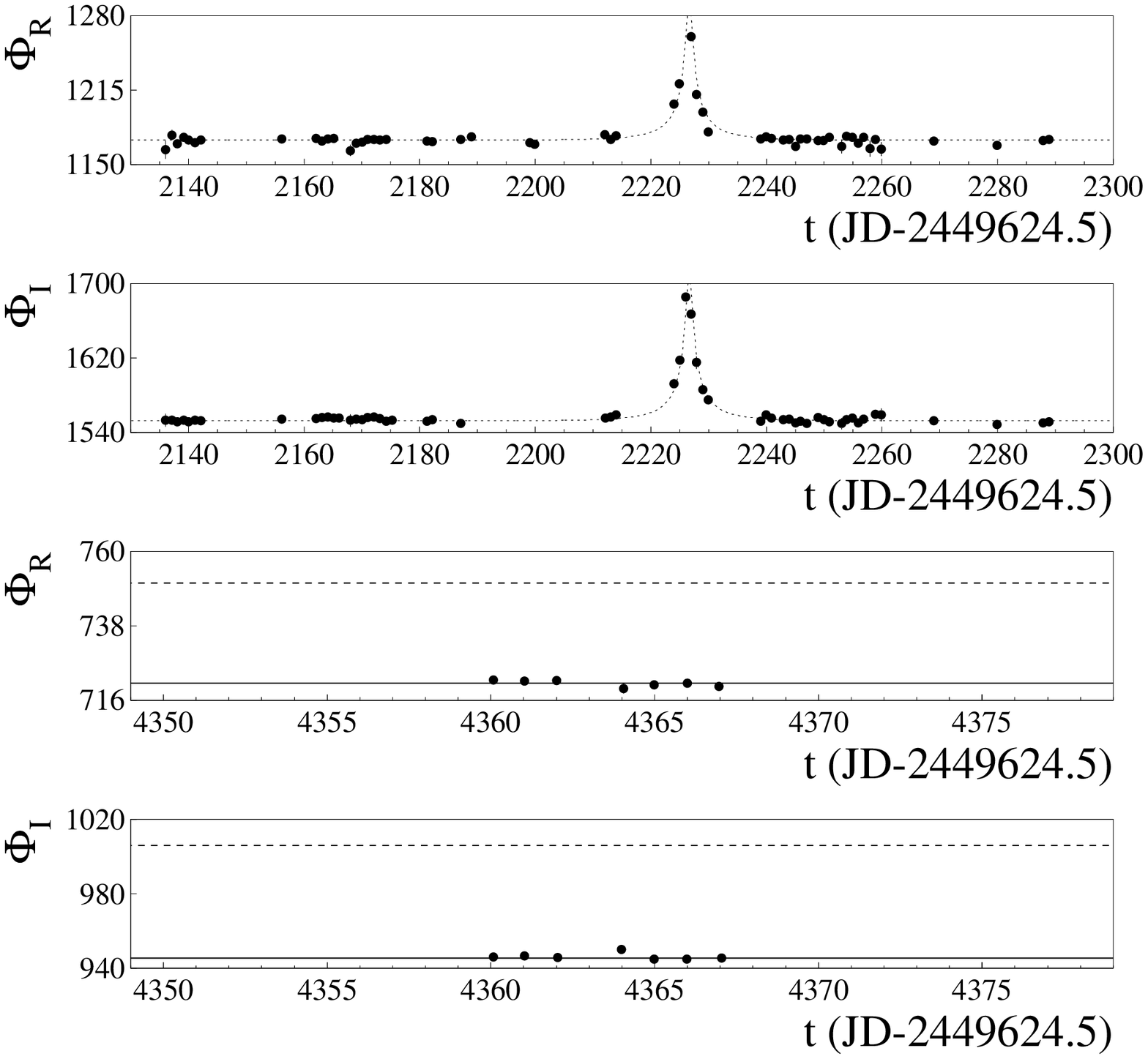}} \caption{
The light curve of the POINT-AGAPE PA-S3 microlensing
candidate \citep{paulin03,novati05} together
with its extension in the OAB data.
In the INT data the dotted line is the best
\cite{pacz86} fit; in the OAB data,
the solid lines indicate the background level, 
while the dotted lines represent the flux deviation corresponding
to the observed flux deviation at maximum for the POINT-AGAPE variation.
The ordinate axis units are flux in ADU s$^{-1}$. 
\label{fig:oab_s3}}
\end{figure}

Since microlensing variations are quite unlikely
to repeat, measuring a constant flux from follow-up observations provides
further evidence that the previously observed signal has indeed been
caused by microlensing.
Our target fields contain three of the six
candidates reported by the POINT-AGAPE collaboration
\citep{novati05}, PA-N1, PA-S3 and PA-S7; beside
PA-N1, two more among the 14 reported by the MEGA collaboration \citep{mega06},
MEGA-ML-3 and MEGA-ML-15; beside PA-S3, the second 
candidate reported by the WeCAPP collaboration, WeCAPP-GL2 \citep{wecapp03}.
\emph{All} of the light curve extensions within our data set 
of the previous variations appear to be \emph{stable},
namely, we do not observe any flux variation beyond the 
background noise level compatible with the observed
microlensing flux variation. As an example, 
in Fig.~\ref{fig:oab_s3} we show the PA-S3 light curve
together with its extension in the OAB data.

\subsection{A Nova like variation} \label{sec:nova}

\begin{figure}
\resizebox{\hsize}{!}{\includegraphics{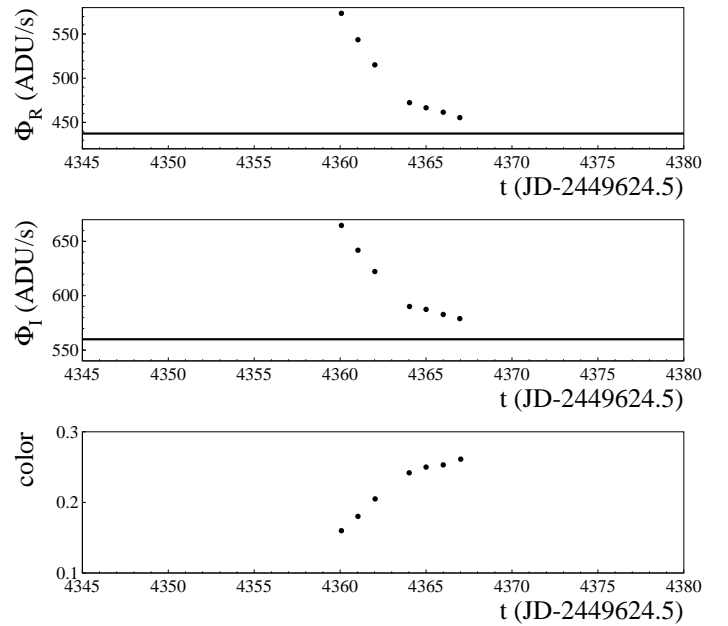}} \caption{
The light curve of the Nova detected within the OAB data.
Top to bottom, $R$, $I$ bands and 
color data (as defined in Fig.~\ref{fig:var}) are shown.
The solid lines ($R$ and $I$ data) indicate the estimate
of the background level. 
\label{fig:nova}}
\end{figure}

Lastly, we discuss the result of a search for very bright flux variations 
($R(\Delta\phi)<19$). One flux variation survives this selection (Fig.~\ref{fig:nova}),
and this appears to be a nova-like variable
(its extension on the POINT-AGAPE data set
appears to be stable) located in RA=$0{\textrm h}42{\textrm m}33{\textrm s}$, 
DEC=$41^\circ 10' 06''$ (J2000).
We estimate the magnitude and color at maximum
to be $R(\Delta\phi)\sim 17.5$ and $R-I\sim -0.1$.
The rate of decline, about 2 magnitudes during
the 7 nights of our observational period,
puts this nova among the ``very fast'' ones
in the speed classes defined in \cite{warner89}.
The (strong) color evolution is rather unusual,
as it got redder during descent.
In the POINT-AGAPE nova catalogue \citep{novae04},
there was only one such object, PACN-00-07,
showing a similar color evolution but
also characterised by a fainter
magnitude at maximum and a slower speed of descent.

The issue of the expected nov\ae~ rate in M31 is still a matter of debate. \cite{novae06}
evaluate a rate of $\sim 38$ ($\sim 27$) nov\ae/years for the bulge
(disc) respectively, while previous works
pointed to somewhat smaller values (e.g. \citealt{capaccioli89,shafter01}).
Our detection of 1 nova during an overall period of 11 days
is in any case in good agreement with these expectations (restricted
to the bulge region only and using the first estimate,
we derive an expected number of nov\ae~of $\sim 1.1$).

\section{The expected microlensing signal} \label{sec:ml}

To predict the number and characteristics
of the expected microlensing signal for the different 
lens populations (Galactic halo and components of M31), we need
both an astrophysical model for all the physical quantities that
determine the microlensing events
(which includes brightness profile, spatial mass density,
velocity distributions, luminosity function for the sources, and
mass spectrum for the lenses)
and a model reproducing both the observational setup
and the selection pipeline.
Because of the huge parameter space
involved, we use  a Monte Carlo simulation to carry out this program.
In particular, we make use of the
simulation described in \cite{novati05}, adapted to
the OAB observational setup.

\begin{figure}
\resizebox{\hsize}{!}{\includegraphics{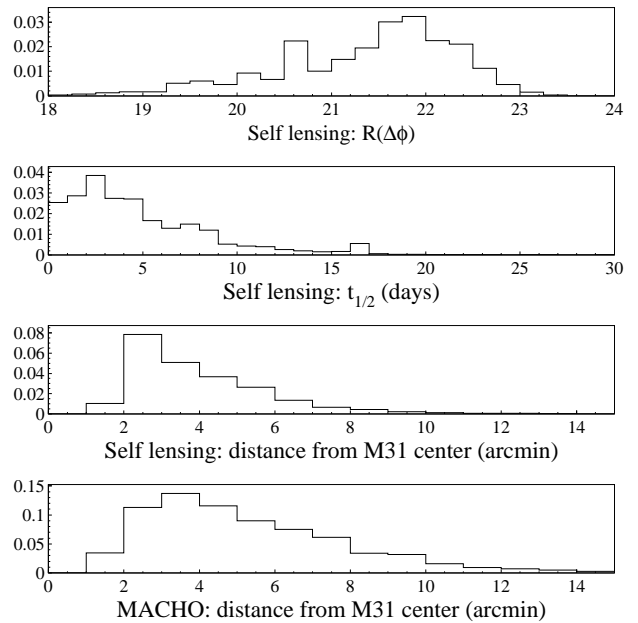}} \caption{
Results of the Monte Carlo simulation:
from top to bottom we show the histograms of the expected flux deviation
at maximum, duration distribution
for self-lensing events, and the distance (from the M31 center)
distributions for self-lensing and for MACHOs. The units on the ordinate 
axes are the number of events.
\label{fig:mc}}
\end{figure}

In Fig.~\ref{fig:mc} we report the results, obtained using the fiducial astrophysical model
discussed in \cite{novati05}, for the flux deviation at maximum
and duration distributions expected for self-lensing events
(the corresponding distributions for 0.5 M$_\odot$ MACHOs
are almost indistinguishable)
and, for both self lensing and MACHOs,
the expected distance from the M31 center
distribution. In particular we recover the well known results that
most of the microlensing events are expected to last only a few days.
We also stress the difference, already apparent within our relatively small
field of view, between the spatial distributions due to luminous and
MACHO lenses.   The latter is much broader, implying that this
diagnostic can be used to distinguish between the two populations.
Note that here we are considering the distance-from-the-M31-center
statistics rather than the expected asymmetry in the spatial
distribution of M31 halo lenses \citep{crotts92,jetzer94}.
The M31-center-distance statistic is sensitive to the different
mass distributions of stars and dark matter, although it is
a zeroth order approximation since it ignores the additional
difference due to the expected asymmetry of the microlensing signal.
We adopt this zeroth-order approach because
the refinement needed to include the asymmetry information
would require substantial additional analysis:
as was pointed out by \cite{an04}, the study of variable stars
demonstrates that differential extinction
could induce a similar asymmetric signal on the spatial distribution
of self-lensing events.
Note also that our choice for the field position
has not been chosen in order to optimise such an analysis,
but rather to maximise the overlap
with the fields of previous campaigns.
Let us note that a real ``second generation'' pixel lensing experiment
should cover a much larger field of view than ours,
both to increase, for a given time baseline,
the expected rate of events but especially in order to better disentangle
the self-lensing signal from the MACHO signal. 

To estimate the \emph{number} of expected events, we reproduce the actual sampling
of this pilot season and implement a basic selection for microlensing events
(asking for the presence of a significant bump),
and take into account the results of the analysis carried out
in Sect. \ref{sec:var} by restricting to the subsample  of $R(\Delta\phi)<22$ variations.
As a results, we predict  $\sim 0.17$ self-lensing events
and $\sim 0.54$ MACHOs (for full M31 and Galactic
halos with 0.5 M$_\odot$ MACHO objects). As discussed in \cite{novati05},
the predictions of the Monte Carlo simulation are quantifiable as ``over-optimistic'',
so that these figures, for the given astrophysical model,  should be taken as an \emph{upper}
limit to the actual number of expected events because we have not factored in the
efficiency of the pipeline.

In order to increase the available statistics
we need a longer time baseline. An aspect here deserves
to be stressed. As the expected duration of the events we are looking for is of the same
order of the length of our present baseline, because of ``boundary'' effects,
the number of expected events should increase
more than linearly with the overall baseline length
(of course, this is no longer true as soon as the baseline is long enough).
This holds under the condition that no gaps are introduced into the sampling,
clearly showing the importance of an appropriate observational strategy. 
For example, for a full two-month campaign we 
predict, again for the sub sample of $R(\Delta\phi)<22$ variations,
$\sim 1.3$ (5.1) self-lensing (0.5 M$_\odot$ MACHO) events, respectively.
Finally, we note that we have obtained similar results
within a parallel analysis carried out following 
the approach outlined in \cite{gin06a,gin06b}.

As for the astrophysical model, we recall that \cite{mega06},
using a different model for the luminous components of M31,
obtained a significantly higher expected contribution of the self-lensing signal
relative to that
evaluated with the fiducial model discussed
in \cite{novati05}, which we are also using in the present analysis. 
Hence, the full-fledged campaign that we are planning will be
important for understanding the disputed issue of M31 self-lensing
as well as MACHO dark matter.

\section{Conclusions} \label{sec:end}

Based on pilot season observations of M31 during 11 consecutive nights in
September 2006 and a Monte Carlo simulation for the expected properties of
microlensing events caused by lenses in the Galactic halo or M31,
respectively, we have shown the feasibility of an extended campaign
with the 1.5m Loiano telescope being able to resolve the current puzzle of
the origin of microlensing events involving extragalactic sources.

In particular, we were able to identify known variable stars from our data
thanks to the tight sampling and despite the short time range covered.
Reported microlensing candidates within our field of view have shown no
further variation, therefore the microlensing interpretation was
confirmed. Moreover, a nova variable showed up in our data.

As for the microlensing signal, we have stressed the importance
of an appropriate sampling for the observations,
and discussed the results of a Monte Carlo
simulation of the present experiment. In particular, we have shown
how the expected spatial distribution
for self-lensing and MACHO events can allow us
to disentangle the two contributions. Finally, we have provided an evaluation
of the expected number of microlensing events for the present pilot season
and discussed quantitatively the possible output
of a longer baseline campaign.

\begin{acknowledgements}
The observational campaign has been possible
thanks to the generous allocation of telescope
time by the TAC of the Bologna Observatory and
to the invaluable help of the technical staff.
In particular, we thank Ivan Bruni for accurate and precious
assistance during the observations.
We thank the POINT-AGAPE collaboration for access to their database.
\end{acknowledgements}

\bibliographystyle{aa}
\bibliography{biblio}
\end{document}